\documentclass{article}
\usepackage{graphicx}

\begin{document}

\title{How the conservation of charge can lead to a faster-than-\textit{c}
effect: A simple example}
\date{White Paper of December 3, 2010 by R. Y. Chiao}
\author{}
\maketitle
\begin{figure}
\includegraphics[width=6in]{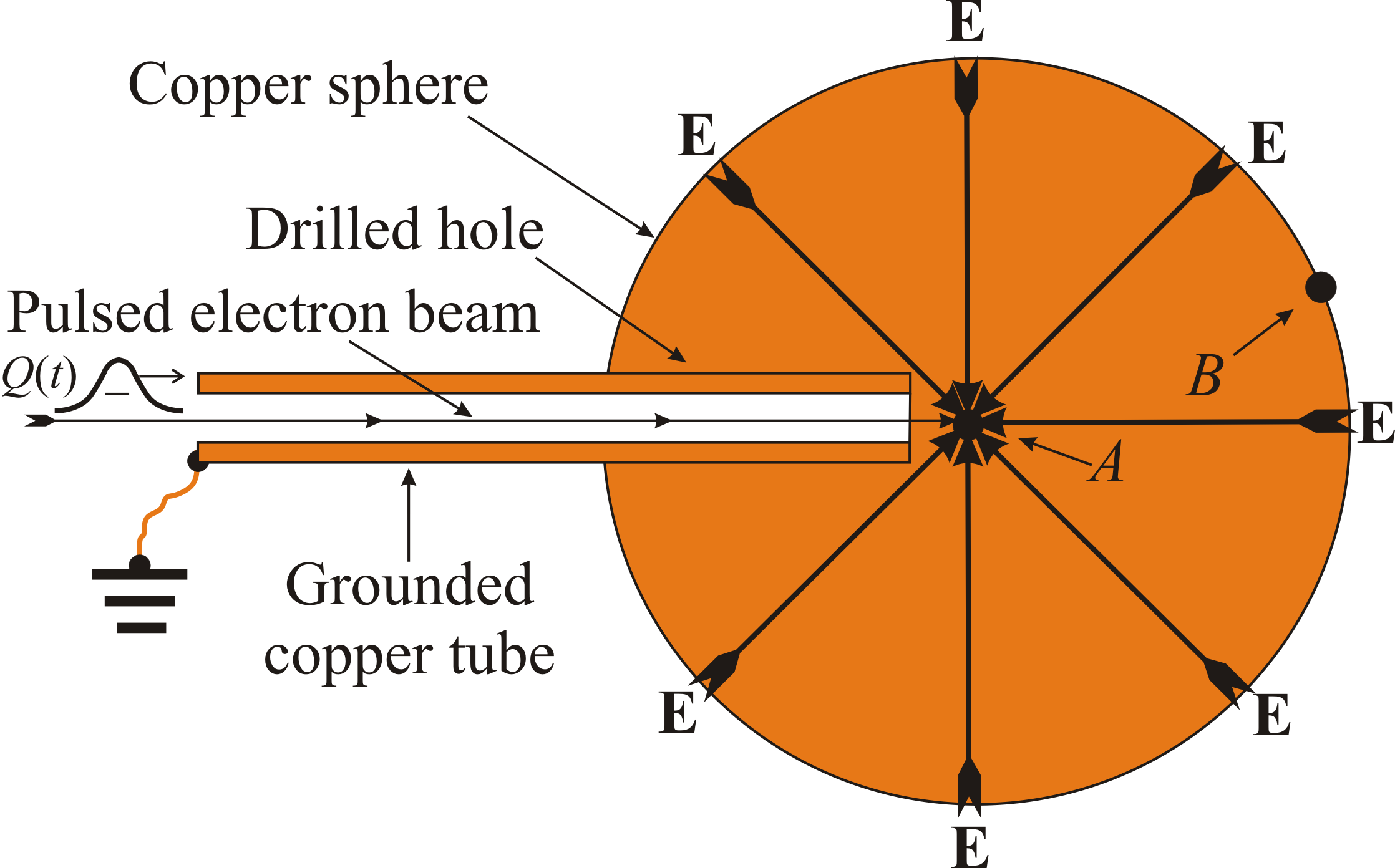}
\label{copper-sphere}
\caption{A pulsed electron beam enters
through a grounded copper tube inserted into a drilled hole that ends just
before the center of a copper sphere. The tube fits tightly into the hole.
The beam is stopped at the center of the sphere at point $A$, and its charge
is deposited there. This deposited charge $Q(t)$ produces radial lines of electric
field $\mathbf{E}$ inside the sphere. How quickly does some of the deposited
charge at point $A$ re-appear as charge at an arbitrarily chosen point $B$
on the surface of the sphere?}
\end{figure}

Within the general relativity community, there is still much doubt that any
non-trivial faster-than-light effects can ever occur. In particular, it is
widely believed that superluminal mass currents cannot exist in Nature.
However, there now exists a substantial amount of experimental evidence that
superluminal group velocities do indeed exist in Nature. For example, the
tunneling of quantum particles, including photons, through a classically
forbidden barrier has been shown to occur at speeds that exceed $c$ \cite%
{Chiao-Steinberg}.

Thus \emph{quantum} currents have been demonstrated to become superluminal
under certain circumstances, in the sense that, within quantum matter, these
currents have been shown to transport particles faster than $c$.
Superluminal quantum \textit{mass} currents can therefore exist within
matter, and there has been a recent theoretical paper \textquotedblleft
Faster than light?\textquotedblright\ by Geroch \cite{Geroch}, which
suggests that the community should take a serious look at the possibility of
superluminal effects which can occur in standard general relativity, and yet
they do not violate causality. Recently, it has also been suggested that
superluminal mass currents within quantum matter may imply the possibility
of the generation and reflection of gravitational waves quantum mechanically
that would be impossible classically \cite{Prague}.

The purpose of this White Paper is to give a simple example of how the
conservation of charge can lead to superluminal effects. Let us start from
the continuity equation%
\begin{equation}
\nabla \cdot \mathbf{J}+\frac{\partial \rho }{\partial t}=0
\label{continuity equation}
\end{equation}%
where $\mathbf{J}$ is the electrical current density, and $\rho $ is the
electrical charge density in an ordinary conducting body, such as a copper
sphere.

Consider the simple thought experiment illustrated in Figure 1. A pulsed electron beam enters through a long, hollow copper
tube, which is grounded near its entrance so that it forms a Faraday cage
that shields the electrons inside the tube \cite{Faraday cage}. This tube is
inserted into a hole which is drilled into a solid copper sphere in such a
way that the hole stops just before it reaches the center of the sphere at
point $A$. Moreover, the tube forms a tightly fitting sleeve within the
copper sphere, and therefore makes an intimate electrical contact with the
walls of the drilled hole within the sphere.

Hence the electrons within the tube, as they approach the center of the
sphere, cannot \textquotedblleft see\textquotedblright\ the electrons within
the volume of the copper sphere, nor can the electrons within the volume of
the sphere \textquotedblleft see\textquotedblright\ the electrons within the
tube, until the pulsed beam of incident electrons strikes the copper
material just before of the center of the sphere, and these electrons come
to a halt at point $A$. Thus electrical charges will be deposited at the
center of the sphere at time $t=0$.

The Coulomb repulsion between the charges thus deposited at point $A$ will
drive them towards the surface of the sphere. However, there are two
different paths by which these charges can reach the surface: They can
either travel luminally (i.e., at the speed of light) as \textit{surface}
currents along the inner radius of the copper tube, or they can travel
superluminally (i.e., faster than the speed of light, as we shall show
presently) as \textit{volume} currents within the body of the copper sphere
towards the surface.

In the latter case, the charges will be driven towards the surface by the 
\emph{monopole} Coulomb field from the point charge source at $A$, which are
indicated by the \emph{radial} lines of electric field $\mathbf{E}$ in
Figure 1 emanating from point $A$. These $\mathbf{E}$
field lines will then produce, via Ohm's law, a current density $\mathbf{J}$
of electrons within the volume of the copper sphere that\ will flow out
radially towards the surface of the sphere.

However, if the inner radius $a$ of the copper tube is chosen to be
sufficiently small, there will exist a sufficiently high cutoff frequency of
the fundamental TM$_{01}$ mode of propagation of EM waves within the tube
viewed as a waveguide, such that, if the spectrum of frequencies in the incident
Gaussian charge pulse lies well below this cutoff frequency, the luminal
path for the charges to flow along the inner surface of the tube will be
exponentially suppressed by this waveguide cutoff, effectively leaving only
the superluminal path for the charges to flow through the volume of the
sphere.

Furthermore, charges originating from point $A$ that reach the surface via
the volume of the sphere can only reach the connection to ground at the far
left end of the copper tube by traveling as luminal charges propagating
along the outer surface of the section of the tube that sticks out from the
surface of the sphere in Figure 1. Thus there could exist
a considerable amount of dwell time on the surface of the sphere for the
superluminal volume charges, before these charges could leak into the
ground, since they would have to travel as luminal charges along the outer
surface of an arbitarily long section of the tube, which extends all the way
from the surface of the sphere to the tube's entrance at the far left end,
where the connection to ground is made.

Let us assume that Ohm's law holds at every point inside the volume of the
sphere, i.e.,%
\begin{equation}
\mathbf{J}=\sigma \mathbf{E}  \label{Ohm's law}
\end{equation}%
where $\mathbf{J}$ is the electrical current density, $\sigma $ is the electrical conductivity, and $\mathbf{E}$\ is the
electric field. This law is valid at the classical, macroscopic level of
description. Substituting (\ref{Ohm's law}) into the continuity
equation, we obtain%
\begin{equation}
\nabla \cdot \left( \sigma \mathbf{E}\right) +\frac{\partial \rho }{\partial
t}=\sigma \nabla \cdot \mathbf{E}+\frac{\partial \rho }{\partial t}=\sigma 
\frac{\rho }{\varepsilon _{o}}+\frac{\partial \rho }{\partial t}=0 ,
\end{equation}%
where we have used the Maxwell equation%
\begin{equation}
\nabla \cdot \mathbf{E=}\frac{\rho }{\varepsilon _{o}}{ ,}
\end{equation}%
and where we have assumed that the conductor is a homogeneous body, so that
the conductivity $\sigma $ is independent of position within the conductor. 

There results a first-order, homogeneous, linear partial differential
equation in time%
\begin{equation}
\frac{\partial \rho }{\partial t}+\frac{\sigma }{\varepsilon _{0}}\rho =0
\label{1st order PDE}
\end{equation}%
or, equivalently,%
\begin{equation}
\frac{\partial \rho }{\partial t}=-\frac{\sigma }{\varepsilon _{0}}\rho =-%
\frac{1}{\tau }\rho { .}
\end{equation}%
Given that $\sigma \approx 6\times 10^{7}$ S m$^{-1}$ for copper, one
obtains the time scale $\tau $ for the rate of disappearance of the charge
at any point inside the sphere where $\rho $ is initially nonzero, viz.,%
\begin{equation}
\tau =\frac{\varepsilon _{0}}{\sigma }{ }\approx 1.4\times 10^{-19}%
{\rm { sec,}}
\end{equation}%
which is approximately the time that it takes light to traverse a Bohr
radius.

The physical significance of $\tau $ is that it denotes the time scale for
the disappearance of the charge deposited at the center of the sphere, and
for its reappearance on the surface of the sphere. Note that this time scale
is independent of the distance separating points $A$ and $B$.

The general solution to the partial differential equation (\ref{1st order
PDE}) at an interior point $\mathbf{r}$ within the sphere at time $t$ is%
\begin{equation}
\rho (\mathbf{r},t)=\rho _{0}(\mathbf{r})e^{-t/\tau }
\label{general solution for rho}
\end{equation}%
where $\rho _{0}(\mathbf{r})$ is to be determined by the initial conditions
inside the copper sphere.

Letting the time $t$ approach the distant past by taking the asymptotic
limit $t\rightarrow -\infty $ in (\ref{general solution for rho}), one
determines at an arbitrary time $t$ in the distant past when the exponential
factor $\exp (-t/\tau )$ is at least as large as some large number, $M$, say, 
and at every point $\mathbf{r}$ in the interior of the
sphere where charge is not being deposited, that 
\begin{equation}
\lim_{t\rightarrow -\infty }\rho (\mathbf{r},t)=\lim_{t\rightarrow -\infty
}\rho _{0}(\mathbf{r})e^{-t/\tau }=\rho _{0}(\mathbf{r})\lim_{t\rightarrow
-\infty }e^{-t/\tau }\geq\rho _{0}(\mathbf{r})M =0{ .}
\end{equation}%
Solving for $\rho _{0}(\mathbf{r})$, one finds that 
\begin{equation}
\rho _{0}(\mathbf{r})=0
\end{equation}%
for all $\mathbf{r}$
except at the center and the surface of the sphere. This is because in the distant past, as one takes the limit $t\rightarrow
-\infty $, i.e., long before the arrival of the main peak of the charge
pulse --- which we shall assume for simplicity to be a Gaussian pulse whose
peak arrives at $t=0$ --- the sphere is initially electrically neutral
everywhere in its interior, so that one requires that $\rho _{0}(\mathbf{r}%
)=0$ for all interior points $\mathbf{r}$. This implies that, at all later
times, and for all interior points $\mathbf{r}$, except for point $A$ at the
center of the sphere, where the charge of electron beam is being deposited,
and except for the points $B$ on the surface of the sphere, where the
deposited charge can reappear,%
\begin{equation}
\rho (\mathbf{r},t)=\rho _{0}(\mathbf{r})e^{-t/\tau }=0{\rm  ,}
\end{equation}%
including when the main peak of the Gaussian pulse arrives at $t=0$.

Therefore, no charge can accumulate at any point $\mathbf{r}$ in the
interior of the sphere, other than at the point $A$ at the center and at
points $B$ on the surface. This is true at \emph{all} times, including the
time when the maximum of the charged pulse occurs as the incident electron beam
reaches the copper sphere at point $A$. Physically, this arises from the
fact that the lines of flow of the current density $\mathbf{J}$ cannot terminate at
any interior point $\mathbf{r}$, because, in order for the body to stay
neutral, the positive charge of the ionic lattice of the copper material must
always be \textit{exactly} compensated by the negative charge of the electrons
at all interior points at all times, including the moment when the incoming
charge, which is being deposited at the center by the beam, reaches a
maximum.

Hence the only places within the sphere where charge can accumulate, and
therefore where the charge density can change with time, is either at point $%
A$ at the center of the sphere, where the charge from the incident Gaussian
pulse is being deposited, or at points on the surface of the sphere, such as
the arbitrarily chosen point $B$, where some of this deposited charge can
re-appear. At each instant of time, however, the \textit{total} charge of
the entire system must always be \textit{exactly} conserved.

This implies that at the quantum, microscopic level of description, 
the disappearance of an individual electron, such as at point $A$,
must always be accompanied by its \textit{simultaneous} reappearance at an
arbitrarily far-away point on the surface, such as at point $B$, at \textit{%
exactly} the same instant of time \cite{Simultaneity}. Otherwise, the
principle of charge conservation would be violated at the quantum level of
description of individual events.

We shall call this counter-intuitive effect \textquotedblleft instantaneous
superluminality due to charge conservation.\textquotedblright\ Note that
this superluminal effect does not violate relativistic causality because the
incident charge pulse can have an \emph{analytic} waveform, e.g., a
Gaussian, with a finite bandwidth (i.e., with frequencies less than the
plasma frequency of the metal). There exists no discontinuous
\textquotedblleft front\textquotedblright\ within a Gaussian waveform,
before which the waveform is \textit{exactly} zero \cite{Front velocity}.
Such a \textquotedblleft front\textquotedblright\ would necessarily contain
infinitely high frequency components that would exceed the plasma frequency,
at which point the concept of electrical conductivity $\sigma $, and
therefore the above analysis, would break down.

Thus this \textquotedblleft instantaneously superluminal\textquotedblright\
effect has similarities with that of a Gaussian wavepacket tunneling through
a tunnel barrier in quantum mechanics, whose early analytic tail contains
all the information needed to reconstruct the entire transmitted wavepacket,
including its peak, earlier in time \textit{before} the incident peak could
have arrived at a detector traveling at the speed of light \cite%
{Chiao-Steinberg}.

To sum up, charge conservation leads to situations in which causal,
faster-than-$c$ effects can occur. At the quantum level of description, such
effects can lead to causally superluminal charge and mass currents inside
matter.


\begin{thebibliography}{9}
\bibitem{Electron wavelength} The electrons in the incoming electron beam
have a short DeBroglie wavelength, and are not affected by this
waveguide-type cutoff for EM waves.
\bibitem{Front velocity} As has been pointed out by Sommerfeld \cite%
{Chiao-Steinberg}, causality in special relativity is represented by the
statement that, inside all matter, the \emph{front} velocity does not exceed 
$c$, and not by the statements that the \emph{group} and the \emph{phase}
velocities do not exceed $c$.
\end{thebibliography}
\end{document}